\documentclass[twocolumn,showpacs,preprintnumbers,amsmath,amssymb]{revtex4}
\usepackage{graphicx}
\usepackage{dcolumn}
\usepackage{bm}

\begin{document}

\title{Coulomb screening and collective excitations in a graphene bilayer}
\author{Xue-Feng Wang and Tapash Chakraborty}
\affiliation{Department of Physics and Astronomy,
The University of
Manitoba, Winnipeg, Canada, R3T 2N2}

\begin{abstract}
We have investigated the Coulomb screening properties and collective
excitations in a graphene bilayer. The static screening effect is
anisotropic and is much stronger in the undoped graphene bilayer
than in a monolayer graphene \cite{ando1}. The dynamic screening
shows the properties of a Dirac gas in the low frequency and that of a
Fermi gas in the high frequency. The transition from the Dirac to the
Fermi gas is also observed in the plasmon spectrum. Finally, we find
that an electron gas in a doped graphene bilayer has quite similar
properties as those of a Fermi gas in materials containing two
energy valleys.
\end{abstract}
\maketitle

The realization of the monolayer and the bilayer graphene -- the
two-dimensional crystal of one and two layers of carbon atoms, has
opened a new door for exploration of the fundamental physics and
also fabrication of nanoelectronic devices \cite{geim,neto}. Being
different from the multilayer graphene or graphite, the monolayer
and the bilayer graphene are intrinsically zero-gap semiconductors.
The monolayer graphene has the
properties of a chiral Dirac gas while the graphene bilayer has
the energy band of a chiral Fermi gas at high energies with several
quasi-Dirac points at the bottom of the band \cite{mcca,kosh,nils,ohta}.
Consequently, a comparison of the physical properties between them
would offer new understanding and provide interesting predictions about
the different behaviors between the charged chiral-Dirac and the
Fermi gases.

In normal two-dimensional semiconductors, the many-body effects on the
Coulomb interaction has been extensively studied using different methods.
One of the most successful and widely accepted approaches is the
random-phase approximation (RPA). In this approximation it is assumed
that only the single-particle excitations of the same wavevector as the
Coulomb interaction plays an effective role in the screening process
while the effects of others having different wavevectors cancel out.
In the past years, there have been a few studies reported in the
literature on the Coulomb screening and the collective excitation
spectrum in the monolayer graphene using the RPA \cite{ando2,khve,wang,vafe,hwan}.
In this paper, we present some of the interesting properties of the Coulomb
screening and the collective excitations in a graphene bilayer.

\begin{figure}
\begin{center}
\begin{picture}(250,190)
\put(-0,-35){\includegraphics{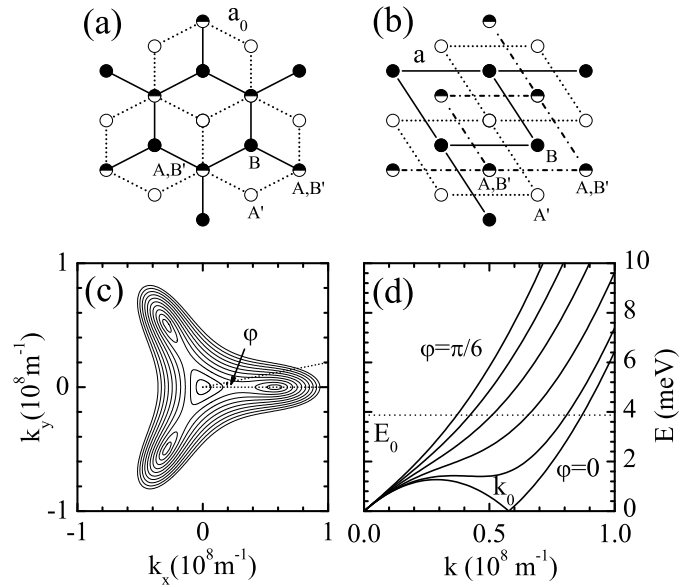}}
\end{picture}
\vspace{1cm} \protect\caption{(a) The hexagonal lattice of the top
graphene layer (solid line) and the bottom layer (dotted line). (b)
Periodic sublattices $A$ and $B^\prime$ (dash-dotted lines), $B$ (solid
lines), and $A^\prime$ (dotted). In (a) and (b), the atoms of the
sublattice $A$ ($B^\prime$), $B$, and $A^\prime$ are denoted by
semi-filled, filled, and empty circles respectively. (c) The contour
lines of the energy in the $k_x-k_y$ plane near the $K$ point. The
corresponding energies, starting with the innermost curve are 0.1$E_0$
to $E_0$ with an increment of 0.1$E_0$. (d) The energy spectrum for
equally separated $\varphi$ from 0 to $\pi/6$
(curves with increasing energy).}
\label{fig:fig1}
\end{center}
\end{figure}

The graphene bilayer is formed by stacking two graphene
layers in the same way as the stacking occurs in graphite,
i.e., the Bernal stacking \cite{mcca,kosh,nils,ohta}. Each graphene layer has
a hexagonal (honeycomb) carbon lattice [Fig.~\ref{fig:fig1}(a)] which
is composed of two periodic sublattices, $A$ and $B$ [Fig.~\ref{fig:fig1}(b)].
In other words, there are two inequivalent lattice sites with atoms
$A$ and $B$ in each unit cell of the periodic lattice. The two sublattices are
displaced from each other along an edge of the hexagons by a distance of
$a_0=1.42$ \AA. In a graphene bilayer, there are four inequivalent sites in
each unit cell, with atoms $A$ and $B$ at the top and $A^\prime$ and $B^\prime$
at the bottom. In the case of the Bernal stacking, the two graphene layers
are arranged in such a way that the $A$ sublattice is exactly on top
of the sublattice $B^\prime$ with a vertical separation of $b_0 = 4$ \AA \
\cite{tric} as shown in Fig.~\ref{fig:fig1}(a) and (b). The system can be
described by the tight-binding model \cite{kosh} characterized by
three coupling parameters, $\gamma_0= 3.16$ eV between atoms $A$ and $B$ or
$A^\prime$ and $B^\prime$ (intra-layer coupling), $\gamma_1=0.39$ eV between
$A$ and $B^\prime$ (the direct inter-layer coupling), and $\gamma_3=0.315$ eV
between $A^\prime$ and $B$, $A$ and $A^\prime$, or $B$ and $B^\prime$ (the
indirect inter-layer coupling).

In the $k$ space, the graphene bilayer has the same hexagonal Brillouin
zone as that of a monolayer graphene. Its physical properties
are mainly determined by the energy spectrum and the wavefunction near
the two inequivalent corners of the Brillouin zone $K$ and $K^\prime$, where
the $\pi^\ast$ conduction band and $\pi$ valence band meet at the Fermi
surface \cite{wallace}. Due to the strong interlayer coupling (the
$\pi$ orbit overlap) both the conduction band and the valence band in a
bilayer are split by an energy of $\sim0.4$ eV near the $K$ and
$K^\prime$ valleys \cite{tric,yosh,ohta}. Since this energy splitting is larger
than the energy range we are interested in from the bottom of the energy
band, we take into account only the upper valence band and the lower
conduction band. The bilayer graphene cannot be
treated as two independent monolayer graphenes with the interlayer coupling
as a perturbation because of the strong interlayer overlap of the $\pi$ orbits.
In contrast, the perturbation treatment of the interlayer coupling is
valid for a normal double quantum well system \cite{wang1} or
in an intercalated graphite \cite{shun}.

In the effective-mass approximation, the electrons in the $K$ valley is
described by a Hamiltonian with a mixture of the linear and the quadratic
terms of $k$ \cite{mcca,kosh}
\begin{equation}
H_K=\frac{\hbar^2}{2m^\ast}
\left(
\begin{array}{cc}
0& k_-^2\\
k_+^2 & 0\\
\end{array}
\right)
-\frac{\hbar^2k_0}{2m^\ast}
\left(
\begin{array}{cc}
0& k_+\\
k_- & 0\\
\end{array}
\right)
\end{equation}
with $k_\pm=k_x\pm k_y$ and ${\bm k}=(k_x,k_y)$ being measured
from the $K$ point. The effective mass of the quadratic term is
$m^{\ast}=2\hbar^2\gamma_1/(3a_0\gamma_0)^2 \approx 0.033$
and the `light' velocity of the linear term is
$v_0=\hbar k_0/2 m^{\ast}=3a_0\gamma_3/2\hbar
\approx 10^5$ m/s with $k_0\approx 10^8/\sqrt{3}$ m$^{-1}$.

The eigenfunction of the above Hamiltonian is $\Psi_{\bm{k}}^\lambda(
\bm{r})=\frac{e^{i\bm{k}\cdot\bm{r}}}{\sqrt{2}}
{\small\left( \array{c}
e^{i\phi_{\bm{k}}}\\
\lambda
\endarray
\right)}$ with the energy
$E_{\bm{k}}^\lambda=\lambda\hbar^2 k\sqrt{k^2-2k_0k\cos3\varphi+k_0^2}/2m{^\ast}$
and the pseudospin angle $\lambda\phi_{\bm{k}}$. Here $\varphi=\arg(k_+)$,
$\phi=\arg(ke^{-2i\varphi}-k_0e^{i\varphi})$ with $\arg(z)$ being the
argument $\theta$ of a complex $z=|z|e^\theta$, and $\lambda=1 (-1)$
for the conduction (valence) band.

For $k\gg k_0$ the electron states are chiral with $\phi=-2\varphi$ and
have an approximately isotropic parabolic energy dispersion
$E_{\bm{k}}^\lambda=\lambda\hbar^2k^2/2m^\ast$. Near $k=k_0$, the energy
dispersion becomes highly anisotropic as shown in Fig.~\ref{fig:fig1}(c)
and (d). The corresponding characteristic energy is
$E_0=k_0^2/2m^\ast=3.9$ meV.
At $E=0$, where the Fermi energy is located in undoped graphene bilayer,
there are four contact points between the conduction
and valence bands: One at $k=0$, the center of the valley and three
satellites at $k=k_0$ in the directions of $\varphi=0$, $\pi/3$, and
$2\pi/3$. They can be treated as four quasi-Dirac points because the
electronic states near each point has a linear energy dispersion and
has the same chirality as those near a Dirac point in the monolayer
graphene. However, compared to the monolayer grahene, the energy
dispersion here is anisotropic and the `light' velocity is about
ten times slower. As illustrated in Fig.~\ref{fig:fig1}(c) and (d),
there is an energy pocket with a depth of about $E_0/4$ at each quasi-Dirac
point. The above peculiar characteristics of the graphene bilayer makes
it quite different from the monolayer graphene in their Coulomb screening
properties as described in the following.

Just as for the case of the monolayer graphene \cite{wang} and other SU(2)
spin systems \cite{wang1}, we find that the dielectric matrix of a graphene
bilayer is again a unit matrix multiplied by a dielectric function
\begin{equation}
\hat{\varepsilon}(q,\omega)=1-v_q\hat{\Pi}_0({\bf q},\omega)
\label{dielsg}
\end{equation}
with the bare Coulomb interaction $v_q=e^2/(2\varepsilon_0q)$ and
the electron-hole propagator
\begin{equation}
\hat{\Pi}({\bf q},\omega) =4\sum_{\lambda, \lambda', \bm{k}}
|g_{\bm{k}}^{\lambda,\lambda'}(\bm{q})|^2
\frac{f(E^{\lambda'}_{\bm{k}+\bm{q}})-f(E^\lambda_{\bm{k}})}
{\omega+E^{\lambda'}_{\bm{k}+\bm{q}}-E^\lambda_{\bm{k}}+i\delta}.
\label{propagator}
\end{equation}
The factor four comes from the degenerate two spins and two valleys at $K$
and $K'$, $f(x)$ is the Fermi function, and the vertex factor reads
$|g^{\lambda,\lambda'}_{\bm{k}}(\bm{q})|^2=
[1+\lambda\lambda'\cos(\theta_{\bm{k}}-\theta_{\bm{k}+\bm{q}})]/2$.
Near the central quasi-Dirac point at $k=0$, the intraband backward
scattering and interband vertical Coulomb scattering are forbidden and
$|g^{\lambda,-\lambda}_{\bm{k}}(0)|^2=
|g^{\lambda,\lambda}_{\bm{k}}(-2\bm{k})|^2=0$.
The same rules also hold for the three satellite quasi-Dirac points.
For a large $k$ ($k\gg k_0$),
$|g^{\lambda,-\lambda}_{\bm{k}}(0)|^2=0$ but $|g^{\lambda,\lambda}_{\bm{k}}
(-2\bm{k})|^2=1$, i.e. the intraband backward transition is allowed but
both the interband backward and vertical transitions are forbidden. The
above selection rules together with the energy dispersion of the carriers
in the graphene bilayer indicate that electrons (holes) close to the
bottom (top) of the conduction (valence) band have very different
behaviors from those away from the bottom (top).

Note that in a monolayer graphene the dielectric function
$\hat{\varepsilon}$ is invariant if all the parameters with the energy
unit, $\omega$, $E_F$, and $k_BT$, and with the wavevector unit, $\bm{k}$
and $\bm{q}$, vary proportionally because of the linear energy dispersion
of the Dirac gas \cite{wang}. As a result, the dielectric function and
the plasmon dispersion is uniform for systems with proportional
parameters. In a graphene bilayer, however, this is not true anymore because
of the nonlinear energy dispersion.

\begin{figure}
\begin{center}
\begin{picture}(250,190)
\put(-0,-35){\includegraphics{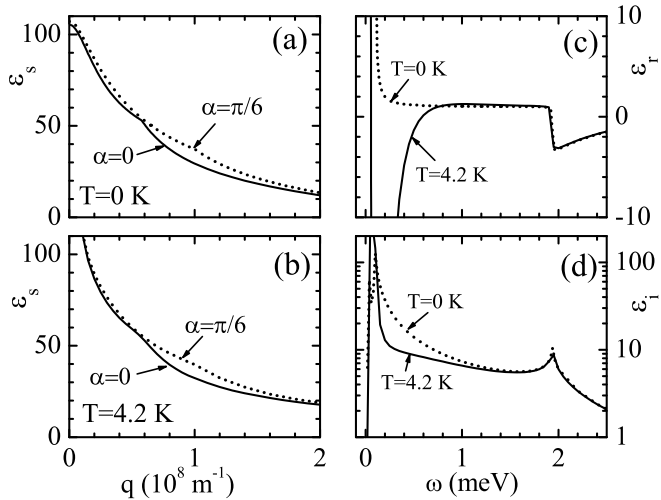}}
\end{picture}
\vspace{1cm} \protect\caption{(a) The static dielectric function
$\varepsilon_s$ vs the wavevector $q$ along the direction $\alpha=0$
(solid) and $\pi/6$ (dotted) at $T=0$. (b) The same as
(a) but at $T=4.2$ K. (c) The real part of the dielectric
function $\varepsilon_r$ vs frequency $\omega$ at $T=0$ (dotted)
and at $T=4.2$ K (solid). (d) The imaginary part of the dielectric
function $\varepsilon_i$ vs $\omega$ at $T=0$ (dotted) and
at $T=4.2$ K (solid). In (c) and (d), $q=0.005\times 10^8$ m$^{-1}$
and $\alpha=0$. In the limit $\omega\rightarrow\infty$, $\varepsilon_r$
gradually approaches to one while $\varepsilon_i$ approaches to zero.}
\label{fig:fig2}
\end{center}
\end{figure}

{\it Coulomb screening}: The static dielectric function at zero
temperature versus $q$ is plotted in Fig.~\ref{fig:fig2}(a). Its
long wavelength limit is given by the properties of the four quasi-%
Dirac points. The central point has an isotropic `light' velocity
$v_0=\hbar k_0/(2m^\ast)=10^5$ m/s while the satellite ones have
the elliptic form of equienergy lines with a minimum `light' velocity
equal to $v_0$ along their radical direction and a maximum of $3v_0$
along the azimuthal direction. The static dielectric constant at $q=0$
is estimated to be $\varepsilon_s=1+3e^2/(8\varepsilon_0\hbar v_0)\approx 105$.
This value is much bigger than the one for the monolayer graphene
($4.5$) \cite{wang}. This means that the long-range Coulomb
interaction is much more strongly screened for the bilayer system,
thanks to a much bigger density of states near the Fermi energy in
a graphene bilayer.

Another characteristic of the bilayer graphene is its screening anisotropy,
especially for scattering at a  distance range of about 10 nm. This is
shown by the difference between the solid and the dotted curves in
Fig.~\ref{fig:fig2}, corresponding to the directions of $\bm{q}$ pointing
to any satellite quasi-Dirac points
($\alpha=0$) or to the middle of any two satellites
($\alpha=\pi/6$) respectively. Here $\alpha$ is the angle between $\bm{q}$
and the $x$-axis. At $q=\sqrt{3}k_0=10^8$ m$^{-1}$, the
wavevector distance between any two satellite quasi-Dirac points,
the anisotropy of $\varepsilon_s$ reaches its maximum whit a
mismatch of 20\% along the different directions.
The shoulder near $q=k_0=0.58\times 10^8$ m$^{-1}$
in the solid curve reflects the strong scattering between the
carriers in the central and the $\varphi=0$ setellite quasi-Dirac
points. At a finite temperature, the energy pockets near the quasi-Dirac points
are partially occupied and the intraband scattering strength is
greatly enhanced. As a result, the static dielectric function near
$q=0$ increases rapidly, as shown in Fig.~\ref{fig:fig2}(b) at $T=4.2$ K.

\begin{figure}
\begin{center}
\begin{picture}(250,190)
\put(-0,-35){\includegraphics{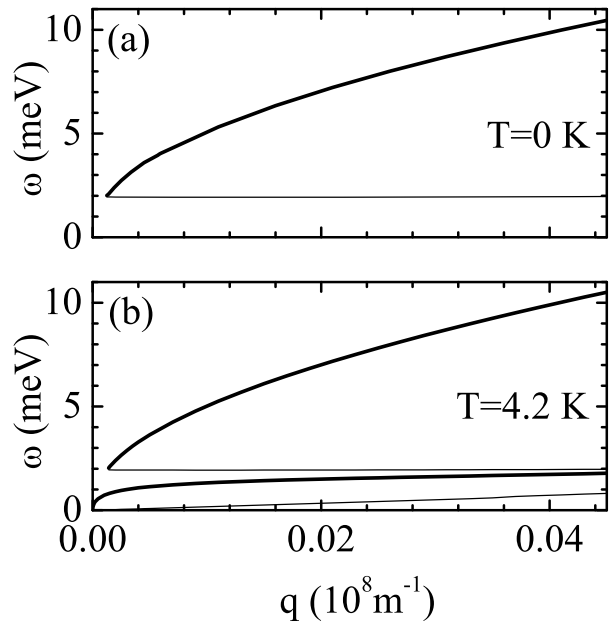}}
\end{picture}
\vspace{2.5cm} \protect\caption{The plasmon spectrum of an undoped
graphene bilayer at $T=0$ (a) and at $T=4.2$ K (b). The thick curves indicate
the weakly Landau damped modes while the thin curves represent the strongly damped
modes.}
\label{fig:fig3}
\end{center}
\end{figure}

In order to have a complete picture of the Coulomb screening and
an insight for understanding the plasmon spectrum in bilayer graphene,
we have calculated the real and imaginary parts of the dynamic
dielectric function versus the energy for a small wavevector $q=0.005$
along $\alpha=0$ at $T=0$ and at $T=4.2$ K in Fig.~\ref{fig:fig2}
(c) and (d).

For $\omega > E_0/2$, the dielectric function of a bilayer graphene
is similar to that of a normal Fermi gas and its temperature dependence
is weak. The step of $\varepsilon_r$ and the peak of $\varepsilon_i$
near $\omega=E_0/2=2$ meV correspond to the single-particle excitation
coupling states with a vanishing group velocity
and having wavevectors located near the middle between the
central and the satellite quasi-Dirac points.
For small $\omega$, however, the dielectric function
becomes more sensitive to the temperature and shows characteristics
of the Dirac gas. One sign of the Dirac gas is the lack of Coulomb screening
($\varepsilon_r\approx 1$) in the energy window between 1 and 2 meV.
Another sign is that a low-energy plasmon mode appears only at a finite
temperature. As shown in Fig.~\ref{fig:fig2}(c), the $\varepsilon_r$ has
no negative value for the energy $\omega < E_0/2$ at $T=0$ but evolves
into a deep negative dip at a finite temperature $T=4.2$ K, when the energy
pockets near the quasi-Dirac points is partially occupied. As a result,
one observes a weakly Landau damped plasmon mode of dispersion
$\omega \sim \sqrt{q}$ at $T=0$ and a couple at finite temperatures.

{\it Collective excitation}: In Fig.~\ref{fig:fig3}(a), we calculate
the plasmon spectrum of an intrinsic bilayer graphene ($E_F=0$). The
dispersion of the weakly Landau damped mode is indicated by the thick
curve and has a $\sqrt{q}$ dependence. Interestingly, the plasmon mode
exists only at the energy higher than $E_0/2$,
i.e., double the depth of the energy pockets in the quasi-Dirac points.
At a finite temperature $T=4.2$ K, another weakly damped plasmon mode
shows up at the energy lower than $E_0/2$
and also has a dispersion of $\sqrt{q}$ near $q=0$,
as illustrated in Fig.~\ref{fig:fig3}(b). The plasmon mode of higher
energy existing at $T=0$ is not sensitive to the temperature.
This temperature dependence of the low and high energy plasmon spectra
represents the marked difference between electron gases having linear (without
collective excitation) and quadratic (with collective excitation)
energy dispersion at $T=0$. We have established earlier that the collective
excitations appear only at a finite temperature for the Dirac gas \cite{wang}.
The electronic states in a graphene bilayer are similar to the Fermi type at the
high energies but reverts to a Dirac type at the low energy.

\begin{figure}
\begin{center}
\begin{picture}(250,190)
\put(-0,-35){\includegraphics{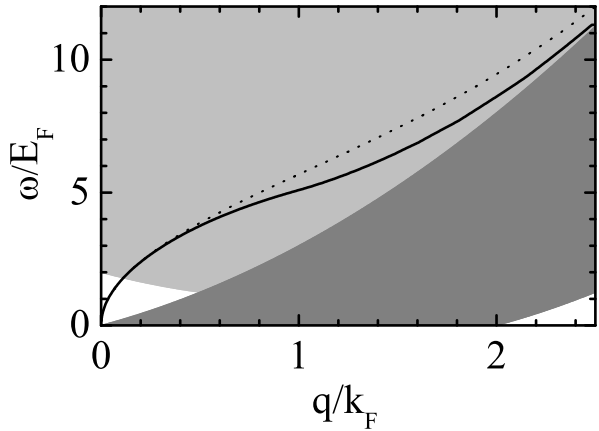}}
\end{picture}
\vspace{-1cm}
\protect\caption{The plasmon spectrum of a doped bilayer graphene
(solid curve) with a typical carrier density of $10^{12}$ cm$^{-2}$.
Correspondingly, $E_F=36.3$ meV and $k_F=1.77\times 10^8$ m$^{-1}$. The
plasmon spectrum in the same system but without chirality is plotted
as a dotted curve for comparison. Intra- (dark shaded) and inter- (light
shaded) band single-particle continuums are also shown.}
\label{fig:fig4}
\end{center}
\end{figure}

The carrier density of the system can be changed by doping \cite{ohta}.
For a typical doping density of $10^{12}$ cm$^{-2}$ \cite{kosh,ohta},
the Fermi energy is high enough from the conduction bottom and
the linear $k$ term in the Hamiltonian can be neglected. The electrons
then have a quadratic dispersion but with chirality and $\phi=-2\varphi$.
Near $q=0$, the plasmon dispersion in a doped bilayer
graphene has again a $\sqrt{q}$
dispersion as shown by the solid curve in Fig.~\ref{fig:fig4}
and shares the same dispersion
$\omega^{2D}_p=[n_e e^2q/2\varepsilon_0 m^\ast]^{1/2}$
with a normal two-dimensional Fermi gas.
To see the effect of the chirality, we plot as a dotted curve the plasmon
dispersion of a normal two-dimensional Fermi gas with two valleys, for
comparison. The two curves overlap for the small $q$ but separate as $q$
increases. The maximum difference in dispersion appears near $q=\sqrt{2}k_F$
when $\bm{k}$ and $\bm{k}+\bm{q}$ form a right angle in the Fermi plane
and the corresponding transition is forbidden in the graphene bilayer due
to its chirality.

In summary, we have explored the Coulomb screening properties and the
collective excitation modes in a graphene bilayer. In an undoped system,
the static dielectric constant is much bigger than that
in a monolayer graphene, due to the existence of four quasi-Dirac
points in each energy valley and the much lower `light' velocity of
the quasi-Dirac points than the one in a monolayer graphene.
The Coulomb screening also shows a strong anisotropy, especially near
the wavevector $q=\sqrt{3}k_0$. The dynamic Coulomb screening has the
characteristics of a Dirac gas
on the low energy side and those of a Fermi gas
on the high energy side. This transition from the Dirac to a Fermi gas is
also reflected in the plasmon dispersion as the wavevector or the energy
increases from zero. In a doped bilayer system, the long wavevector
limit of the collective excitation has the same property as in a
normal two-dimensional Fermi gas with two valleys.

The work has been supported by the Canada Research Chair
Program and a Canadian Foundation for Innovation (CFI) Grant.

\end{document}